\documentclass[sigconf]{acmart}

\AtBeginDocument{%
  \providecommand\BibTeX{{%
    \normalfont B\kern-0.5em{\scshape i\kern-0.25em b}\kern-0.8em\TeX}}}

\copyrightyear{2025}
\acmYear{2025}
\setcopyright{acmcopyright}
\acmConference[CHI '25 Workshop on Tools for Thought]{CHI Conference on Human Factors in Computing Systems}{April 26-May 1, 2025}{Yokohama, Japan}
\acmBooktitle{CHI Conference on Human Factors in Computing Systems (CHI '25), April 26-May 1, 2025, Yokohama, Japan}
\acmISBN{978-1-4503-XXXX-X/18/06}

\usepackage{lipsum}
\usepackage{subcaption}
\usepackage{macros}
\usepackage{float}
\usepackage{wrapfig}

\newcommand{\etal}{et al.}

\begin{document}

\title{Feedforward in Generative AI: Opportunities for a Design Space}

\author{Bryan Min}
\email{bdmin@ucsd.edu}
\affiliation{%
  \institution{University of California San Diego}
  \streetaddress{9500 Gilman Dr}
  \city{La Jolla}
  \state{California}
  \country{USA}
  \postcode{92093}
}

\author{Haijun Xia}
\email{haijunxia@ucsd.edu}
\affiliation{%
  \institution{University of California San Diego}
  \streetaddress{9500 Gilman Dr}
  \city{La Jolla}
  \state{California}
  \country{USA}
  \postcode{92093}
}

\renewcommand{\shortauthors}{Min et al.}

\begin{abstract}

Generative AI (GenAI) models have become more capable than ever at augmenting productivity and cognition across diverse contexts. However, a fundamental challenge remains as users struggle to anticipate what AI will generate. As a result, they must engage in excessive turn-taking with the AI's feedback to clarify their intent, leading to significant cognitive load and time investment. Our goal is to advance the perspective that in order for users to seamlessly leverage the full potential of GenAI systems across various contexts, we must design GenAI systems that not only provide informative feedback but also informative feedforward—designs that tell users what AI will generate before the user submits their prompt. To spark discussion on feedforward in GenAI, we designed diverse instantiations of feedforward across four GenAI applications: conversational UIs, document editors, malleable interfaces, and agent automations, and discussed how these designs can contribute to a more rigorous investigation of a design space and a set of guidelines for feedforward in all GenAI systems.



\end{abstract}

\ccsdesc[500]{Human-centered computing~Interaction paradigms}
\ccsdesc[500]{Human-centered computing~HCI theory, concepts and models}

\keywords{Feedforward, Human-AI Interaction, Generative AI}


\maketitle{}

\section{Introduction}
\label{section:introduction}

\begin{figure*}
    \centering
    \includegraphics[width=1\linewidth]{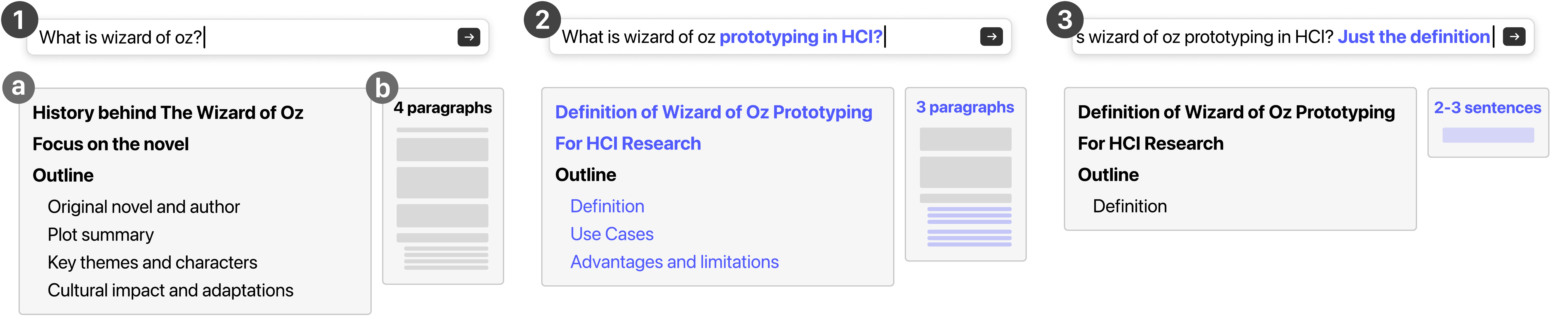}
    \caption{(1) When the user stops typing for a brief moment, the conversational UI presents two feedforward components: (a) a list of key topics and an outline, and (b) a visual minimap of the anticipated length of the response. (2) The user anticipates what the LLM will generate and adjusts their prompt to match it. (3) They also ask for less information, in which the feedforward components update to match the request.}
    \label{fig:feedforward-components}
\end{figure*}

\begin{figure*}
    \centering
    \includegraphics[width=1\linewidth]{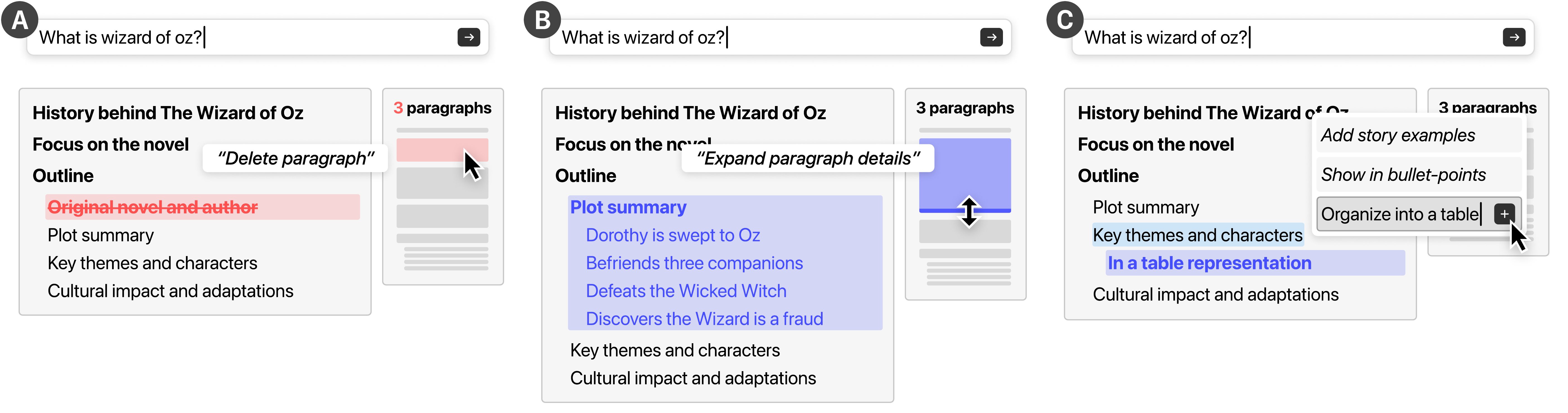}
    \caption{Users can directly interact and manipulate feedforward components by for instance (A) deleting unneeded paragraphs, (B) expanding details about a section, and (C) transforming the representation of the expected content.}
    \label{fig:feedforward-manipulation}
\end{figure*}

Generative AI (GenAI) models are capable of generating a diverse range of content including stories, videos, code, and more. HCI communities have integrated these models into a diverse range of applications to augment learning and teaching \cite{sensecape, jin2024teachAI}, support academic literature review \cite{elicit2024, kang2023synergi, fok2024qlarify, paperweaver}, generate and customize software on the fly \cite{stylette, cao2025jelly, malleableODI, priyan2024dynavis}, and automate web tasks \cite{operator2025, feng2025cocoa}, showcasing their potential to enhance productivity and cognition on a universal scale.

Despite its potential, users inevitably face challenges in trying to anticipate GenAI's responses and aligning their prompts to match their needs.
Subramonyam \etal{} have labeled this set of challenges as the \textit{gulf of envisioning}, which describes the user's gap in identifying and articulating their desired task, providing effective instructions, and anticipating the model’s response \cite{hari2024gulfenvisioning}.
This gap makes it difficult for novices to formulate effective prompts to AI and understand the cause of misalignment \cite{zamfirescu2023johnnyprompt}.

While researchers have explored various approaches to help users better learn and anticipate AI's responses with dedicated spaces for experimenting with prompts \cite{arawjo2024chainforge, angert2023spellburst}, generating multiple outputs in a single space \cite{luminate, gero2024llmsensemakingscale}, asking clarifying questions \cite{deepresearch2025}, and providing widgets to modify generated outputs post-hoc \cite{priyan2024dynavis, masson2024directGPT, graphologue}, these systems primarily focus on offering \textit{feedback}—helping users understand what happened \textit{after} they submit a prompt.
While feedback helps users develop an understanding of the system through exploration, users cannot always anticipate the outcome of their prompts.
Instead, they must rely on iterative trial-and-error, which can incur significant cognitive load and time investment.
If we want users to easily leverage the potential of GenAI systems across diverse contexts, they must be able to anticipate AI outputs without solely relying on feedback.

To achieve this, we aim to advance the perspective that GenAI must not only provide informative feedback but also informative \textit{feedforward}---designs that help users anticipate outcomes \textit{before} performing an action \cite{vermeulen2013feedforward}.
Providing such feedforward in GenAI systems can help users understand what to expect from the AI model before submitting their prompt.
This can reduce excessive conversational exchanges, bridge the abstraction gap between users and AI \cite{liu2023groundedabstraction, hari2024gulfenvisioning, priyan2024imaginingafuture}, and encourage meta-cognition when constructing prompts \cite{tankelevitch2024metacogAI}.

The HCI community has extensively explored feedforward design in various domains, such as gestures \cite{bau2008octopocus}, VR/AR \cite{muresan2023feedforwardVR}, widgets \cite{coppers2019fortunettes, terry2002sideviews}, and proactive systems \cite{allen1999mixedinitiative}.
Furthermore, the community has begun exploring feedforward designs for GenAI, examining trade-offs between the level of detail provided and the cognitive load involved in assessing generated code and images \cite{zhutian2024sketchgenerate, dang2022ganslider}.
However, GenAI's expanding capabilities are shaping an increasingly broad design space of GenAI interfaces.
This rapid growth calls for a more systematic approach to designing effective feedforward across diverse domains---uncovering opportunities for novel interaction techniques, assessing user benefits and challenges across design dimensions, and establishing comprehensive design guidelines.

This paper aims to spark discussion on how we may achieve these goals, enabling users to effectively engage and interact with all forms of GenAI systems.
To lay the groundwork for this discussion, we designed diverse instantiations of feedforward across four GenAI applications: conversational UIs, document editors, malleable interfaces, and agent automations. We then discussed the designs and implementations needed to further develop a comprehensive design space for informative feedforward GenAI systems.

We invite researchers in this community to critically examine past and present GenAI interfaces through the lens of feedforward, draw from feedforward literature in other domains, explore new interaction techniques and frameworks for delivering informative feedforward, and work toward concrete design and implementation strategies applicable to all GenAI systems.

\section{Feedforward in Generative AI}
\label{section:feedforward}

This section aims to explore possible ways to provide feedforward in generative AI systems. We view that effective feedforward in generative AI interfaces should enable users to:
\begin{itemize}
    \item Anticipate what the AI's response or action may be.
    \item Disambiguate their prompt without excessive exchanges in conversation or interaction.
    \item Directly engage with the feedforward content.
    \item  Reflect on their prompting practices with less friction.
\end{itemize}

To explore various ways to design feedforward in GenAI systems, we first explore feedforward in an LLM-powered conversational interface. We then describe scenarios for three other designs of feedforward in GenAI applications.

\begin{figure*}
    \centering
    \includegraphics[width=1\linewidth]{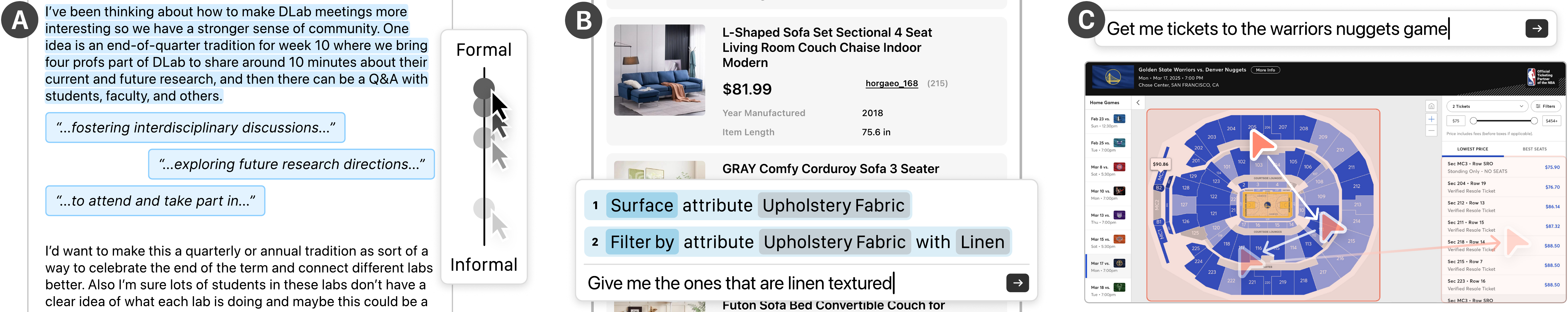}
    \caption{We explored three different applications for designing feedforward in GenAI systems: (A) document editors, (B) malleable interfaces, and (C) agent automation systems.}
    \label{fig:applications}
    \vspace{6pt}
\end{figure*}

\subsection{Feedforward in Conversational LLM UIs}

\subsubsection{Feedforward Components}

Research has shown that users benefit from multiple representations when interacting with LLMs \cite{graphologue, sensecape}.
Therefore, we developed feedforward components that each aim to provide a specific representation that would help users anticipate the LLM's response. 
As the user types a prompt, the conversational UI generates feedforward content inside two feedforward components: 1) a key topic phrase alongside a high-level outline of the anticipated response (Fig. \ref{fig:feedforward-components}.1a) and 2) a minimap view with paragraph blocks (Fig. \ref{fig:feedforward-components}.1b). These representations help users anticipate the topics and subtopics the LLM will generate, as well as whether its response might be too short or too long.


\subsubsection{Manipulating Feedforward Content}

We enable users to manipulate feedforward content in two ways.
First, when users edit their prompt, the feedforward content updates to reflect the changes.
This allows users to identify misalignments between their intent and the feedforward content, encouraging them to add details to the prompt before submitting it.
For instance, they can clarify an ambiguous term (Fig. \ref{fig:feedforward-components}.2) or request a shorter response (Fig. \ref{fig:feedforward-components}.3).
These instant updates to the feedforward components may also spark reflection, such as realizing that it may be useful to understand the advantages and limitations of Wizard of Oz Prototyping rather than just its definition.

Second, users can directly edit content inside the feedforward components. For example, rather than prompting the LLM to remove subtopics from the outline, users can directly delete the subtopic title in the outline component or delete the paragraph in the minimap (Fig. \ref{fig:feedforward-manipulation}A). Users can also drag the edge of the minimap component to expand the feedforward content's level of details (Fig. \ref{fig:feedforward-manipulation}B). This can allow users control ``how far ahead'' they can look into the LLM's response. Additionally, users can highlight parts of the outline, revealing a tooltip that generates a list of possible actions and outputs based on that selection, such as adding examples about a subtopic or organizing the content in a bullet-point list. Users can either select one of these actions or request their own, which will update the feedforward outline (Fig. \ref{fig:feedforward-manipulation}C).

\subsection{Application Scenarios}

Our goal is to explore opportunities for designing feedforward in all GenAI interfaces---beyond conversational UIs.
In this paper, we explore three applications to showcase various feedforward designs across diverse contexts: document editors, malleable interfaces, and agent automations.

\subsubsection{Document Editors}

Developers and researchers have begun to integrate AI into document editors and writing canvases to support various aspects of writing tasks, such as fixing grammar \cite{laban2024inksync}, presenting continuously synchronized outlines \cite{dang2022beyondtextgen}, and providing widgets to adjust the tone of writing along various dimensions \cite{masson2024textoshop, chatgptcanvas2025}.
However, it can be difficult for users to anticipate what the LLM might produce based on their writing and revisions. For instance, moving a slider up to change the tone from ``informal'' to ``formal'' does not indicate exactly how ``formal'' they are making their writing. It is only after submitting the change and reviewing the final result that users realize whether they have achieved the desired outcome. To reduce the amount trial-and-error to achieve the desired result, document editors can provide feedforward that presents example phrases as the user moves the slider up and down, helping the user quickly develop a clear understanding of the level of formality they are adjusting to (Fig. \ref{fig:applications}A). This feedforward mechanism can be expanded to let users control how much detail is shown in each example phrase, allowing them to reveal more or fewer words as needed. This customization helps align the feedforward component with user preferences, reducing cognitive load.




\subsubsection{Malleable Interfaces}

GenAI is additionally enabling software interfaces to become increasingly malleable, enabling users to generate custom functional applications on the fly \cite{litt2025workout, onetwoval2025calculator}, personalized interfaces that blends user activity \cite{cao2025jelly}, and user-defined abstractions in overview-detail interfaces \cite{malleableODI}.
In current generated malleable interfaces, the user prompts for a custom application and provides specifications, and after rounds of clarifying, AI generates and compiles the requested application. However, it can be unclear what kinds of software components the AI might produce based on their prompts alone. For example, if the user asks AI to show only linen-textured couches from a shopping website, the user cannot be sure whether AI will add a filtering operation, perform another search query, or create a brand new list linen-textured couches. Feedforward can present a list of system operations the AI will perform before submitting the prompt, allowing them to quickly assess and revise the operations without unnecessary exchanges with the AI (Fig. \ref{fig:applications}B).




\subsubsection{Agent Automations}

The AI community has also presented demonstrations of AI agents that can automate diverse tasks on the web \cite{operator2025, dia2025}. However, there lacks clear communication of exactly what the agent is doing on the screen. For instance, if a user asks an AI agent to look for tickets to a basketball game, the agent will break the task into steps and perform them by opening and navigating a webpage. During this phase, the user may be unaware of which buttons or components the agent plans to click to find the best prices, which portions of the webpage it will scan for context, or where it will navigate next. Agent automation systems can integrate feedforward visualizations by providing area selections to show where the AI agent is focusing on for context and opaque cursors to tell the user where the agent will click next (Fig. \ref{fig:applications}C). These feedforward representations help users anticipate the agent’s future actions and provide visual affordances to intervene in potential errors by adjusting the selection area or dragging the cursor.

\section{Future Work}
\label{section:discussion}


Our goal is to establish feedforward as a fundamental design component in all GenAI systems. While this paper presents several feedforward designs, we envision a more comprehensive design space to represent and guide feedforward design across all GenAI applications. Based on our four prototypes, we identify three potential design dimensions for GenAI feedforward: representation, level of detail, and manipulability.

First, our examples implemented feedforward representations in the form of outlines, minimaps, lists of operations, example phrases, and multiple cursors. We may identify categories of representations that help distinguish which feedforward representations are more useful for different use cases. For instance, while a list of operations might inform users about the type of UI a GenAI system will generate, a wireframe might better communicate the structure and layout of the UI.

Second, we explored different ways to present varying levels of detail in feedforward. For example, the conversational UI displays an outline summarizing key topics, while the minimap omits textual details and instead presents blocks of paragraphs. This dimension aligns with previous research on feedforward \cite{bau2008octopocus}. A more rigorous investigation could explore optimal levels of detail for different types of feedforward representations in GenAI, as well as allowing users to define the level and type of detail themselves.

Lastly, we explored how users can manipulate feedforward content, either by revising their prompts or by directly resizing, repositioning, or selecting elements. We aim to investigate additional interaction techniques that enhance user control and engagement with feedforward designs.

Future work should expand on this preliminary design space by gathering, analyzing, and critiquing a broader range of GenAI systems to identify variations of feedforward across diverse contexts.

\bibliographystyle{ACM-Reference-Format}
\bibliography{main}

\appendix


\end{document}